\documentclass[prd,reprint,nofootinbib,notitlepage,aps,tightenlines,preprintnumbers,amsmath,amssymb,showpacs,superscriptaddress]{revtex4-1}

\usepackage[hyperindex,breaklinks]{hyperref}
\usepackage{graphicx,epstopdf,amsmath,amsfonts,amssymb,color}

\newcommand{\cm}{\ensuremath{\mathrm{cm}}}

\newcommand{\eV}{\ensuremath{\mathrm{eV}}}
\newcommand{\keV}{\ensuremath{\mathrm{keV}}}

\newcommand{\GeV}{\ensuremath{\mathrm{GeV}}}

\DeclareMathOperator{\binomial}{binom}
\DeclareMathOperator{\gaussian}{gauss}

\begin{document}

\title{New limits on dark photons from solar emission and keV scale dark matter}

\author{Haipeng An}
\affiliation{
Department of Physics, Tsinghua University, Beijing 100084, China}
\affiliation{Center for High Energy Physics, Tsinghua University, Beijing 100084, China}
\author{Maxim Pospelov}
\affiliation{William I. Fine Theoretical Physics Institute, School of Physics and Astronomy, University of Minnesota, Minneapolis, MN 55455, USA}
\author{Josef Pradler}
\affiliation{Institute of High Energy Physics, Austrian Academy of Sciences, Nikolsdorfergasse 18, 1050 Vienna, Austria}

\author{Adam Ritz}
\affiliation{Department of Physics and Astronomy, University of Victoria, 
Victoria, BC V8P 5C2, Canada}

\date{June 2020}

\begin{abstract}
We provide updates to the limits on solar emission of dark photons, or more generally any light vector particle coupled to the electron vector current. The recent 2019 and 2020 electronic recoil data from XENON1T now provides more stringent constraints on these models than stellar energy loss in the sub-keV mass region. We also show that solar emission of dark photons does not provide a good fit to the recent XENON1T excess in the 2-5 keV energy bins. In contrast, the absorption of 2-4 keV mass dark photons that saturate the local dark matter mass density does provide a good fit to the excess, for mixing angles in the range
$\epsilon \in (4-12)\times 10^{-16}$, while satisfying  astrophysical constraints. Similarly, other models utilizing the vector portal can fit the excess, including those with operators that directly couple the dark photon field strength to electron spin.

\end{abstract}
\maketitle

\section{Introduction}

The successful experimental program to scale up the size of underground dark matter detectors based on ultra-pure xenon~\cite{Aprile:2018dbl,Akerib:2013tjd,Cui:2017nnn} has led to a range of significant new constraints on the properties of dark matter. The primary goal of these experiments is the search for thermal relic particles (WIMPs) with weak-scale mass, and the absence of clear signals has resulted in stringent constraints on the sub-weak-scale scattering cross sections with nuclei of any such models. This has in part motivated further attention to the broader landscape of potential dark matter scenarios.

In comparison to underground experiments focused on neutrino detection, that are typically sensitive to energy depositions above a few hundred keV, Xenon-based dark matter detectors provide the leading sensitivity for electronic energy depositions of $~100\,{\rm keV}$ and below. As they are large and very clean ({\em i.e.}~almost free from radioactive contamination and external backgrounds), these experiments are also at the forefront of searches for other exotic particles, beyond the familiar WIMP mass window. For example, light sub-MeV mass dark matter that interacts with 
atomic electrons via absorption or scattering 
\cite{Pospelov:2008jk,Essig:2011nj,An:2014twa,Bloch:2016sjj} is now significantly constrained by recent electronic recoil data from the XENON1T experiment~\cite{Aprile:2019xxb,Aprile:2020tmw}. 
The sensitivity and low background rates 
of Xenon-based experiments has also led to constraints on more energetic sub-components of dark matter in the galactic halo that are created in collisions with energetic 
Standard Model particles. Such examples include light dark matter reflected from energetic particles in the Sun \cite{An:2017ojc,Emken:2017hnp,Zhang:2020nis} or accelerated in interaction with cosmic rays \cite{Bringmann:2018cvk,Ema:2018bih,Dent:2019krz,Cappiello:2019qsw},  or possibly created in the cosmic ray interactions with the atmosphere~\cite{Alvey:2019zaa,Plestid:2020kdm}.

A further application of dark matter direct detection experiments is to search for new light particles emitted from the Sun. While the detection of Standard Model solar neutrinos will have to wait for the next generation of Xenon-based detectors~\cite{Aprile:2015uzo,Akerib:2018dfk}, the current generation of experiments already set meaningful constraints on the emission of light exotic degrees of freedom such as axion-like particles and dark photons \cite{Avignone:1986vm,Pospelov:2008jk,An:2013yua}. While the constraints on axion-like particles are generally 
weaker than those derived from astrophysics and specifically stellar cooling, the limits on dark photons in the
sub 10-eV mass range are competitive with the solar energy loss bounds. 

The goal of this paper is to provide updated bounds on the solar emission of dark photons, as described by the following Lagrangian \cite{Holdom:1985ag}, 
\begin{equation}
\label{L}
 {\cal L}   = {\cal L}_{\rm SM} - \frac{1}{4} (F'_{\mu\nu})^2 + 
 \frac{\epsilon}{2}F'_{\mu\nu}F_{\mu\nu} + \frac{1}{2}m_{A'}^2(A'_\mu)^2.
\end{equation}
Here the primed letters refer to the dark photon, while the un-primed ones refer to the electromagnetic field. The mixing parameter $\epsilon$ is physical unless $m_{A'}$ is strictly zero, in which case it can be rotated away. When $m_{A'}$ is small compared to other dimensionful parameters such as plasma frequency, the stellar energy loss bounds decouple as $\epsilon^2m_{A'}^2$~\cite{An:2013yfc}.
Generalizing models 
with vector particles to other interaction portals, such as gauged $B-L$, also shows relaxed stellar loss bounds (compared to naive expectations) in the limit of small vector mass $m_V$~\cite{Hardy:2016kme}.
 However, if $m_{A'}$ is Higgsed, the Lagrangian~\eqref{L} is amended by
\begin{equation}
 {\cal L}_{\rm higgs} = |(\partial_\mu - i e' A'_\mu)\phi|^2 - V(\phi), 
\end{equation}
and Higgs-strahlung processes into $A' h$ pairs lead to a general non-decoupling of stellar energy loss constraints~\cite{An:2013yua}, controlled by the addional gauge coupling parameter $e'$; $\phi$ is the dark Higgs field with potential $V(\phi)$ and physical excitation $h$.

In the following, we will show that the electronic recoil data from XENON1T~\cite{Aprile:2019xxb}, utilizing only the delayed scintillation signal generated by ionization (S2),
sets new stringent bounds on dark photons that surpass the constraints on $\epsilon$ from XENON10 data~\cite{Angle:2011th} by a factor of a few over a wide range of masses. Given that the absorption signal scales as $\epsilon^4$, this constitutes a very significant improvement. Similarly, the constraints on other vector portal models that utilize an interaction with the electron vector current are also improved. 

The recent data from the XENON1T collaboration~\cite{Aprile:2020tmw}, also adds an intriguing twist, with a low-energy excess in events containing the prompt scintillation signal (S1), between the energy threshold of $\sim 1\,\keV$ and O(5\,{\rm keV}), over the background model. While the most likely explanation for this excess is a statistical fluctuation or unaccounted sources of radioactive contamination, it is nonetheless interesting to explore the models of new physics that may be consistent with such an excess while satisfying other experimental constraints. For example, the absorption of solar axion-like particles cannot be a viable explanation, as 
the astrophysical energy loss constraints are significantly stronger than the 
suggested size of couplings that may lead to an excess. 
In addition, the atomic absorption of light axion-like particles due to the $g_{aee}(2m_e)^{-1}\bar e \gamma_\mu\gamma_5 e \partial_\mu  a$ operator is penalized by a small factor $\omega_a^2/m_e^2$ factor~\cite{Pospelov:2008jk}
making axio-electric cross sections to be additionally suppressed. 
In this work, we show that the absorption of solar dark photons while in principle capable of inducing a large number of events in XENON1T, gives a poor explanation of the observed data for a different reason: the predicted spectrum is considerably softer than is implied by the excess.\footnote{There are other features in the spectrum compared to the collaboration's background estimates. For example, there is a low data point at 17~keV and high data points at 24 and 26~keV, see Fig.~\ref{fig:2} below; we do not address their origin in this work.}

On the other hand, we find that models positing the existence of dark photon dark matter in the keV mass range can fit the excess with $\epsilon$ in the sub-$10^{-15}$ range. (The same conclusion applies to other vector portal models that provide a stable dark matter candidate in this mass range.) Such a small coupling also satisfies the astrophysical stellar cooling bounds. Dark photons are not unique in that respect: an axion-like particle with derivative coupling to electrons, and no direct coupling to photons other than the one radiatively generated at the electron threshold, can fit the data equally well \cite{Takahashi:2020bpq}. (Such models were first introduced  in 2008 in \cite{Pospelov:2008jk}, but the  sensitivity of the direct detection experiments was below the level of stellar cooling constraints at that time.)

The rest of this paper is organized as follows. In the next section we provide an update on solar emission of the
dark photon and related vector portal models in light of the recent XENON1T data. 
In Sec.~\ref{sec:absorption} we address the absorption of light dark matter particles, including dark photon dark matter, in XENON1T and determine the best-fit regions of parameter space consistent with the excess. In Sec.~\ref{sec:discussion}, we reach our conclusions and comment on generalizations of this analysis to the de-excitation of dark matter states.

\section{New bounds on solar dark photons}
\label{sec:newbounds}

The physics leading to the emission and absorption of dark photons has been thoroughly explored in the literature \cite{Redondo:2008aa,An:2013yfc,An:2013yua,Redondo:2013lna,Vinyoles:2015aba}, and is by now well understood. On the emission side, in the limit of $m_{A'} \ll \omega_{p}$,
the longitudinal mode dominates, while in the higher mass range the transverse
mode is typically more important. When the resonant conditions are fulfilled, 
stellar energy loss is dominated by the plasmon-dark photon oscillation. On the detection side, a generalization of the same mechanism applies \cite{An:2013yua}, and knowledge of the dispersive and absorptive part of the refraction index of xenon, as functions of frequency, provides sufficient 
information for the calculation of the dark photon absorption. We refer the reader to the earlier literature for further details. 

Once absorbed in the detector, dark photons create an ionization signal in Xenon~(S2), 
and with sufficient energy release ($E>1$ keVee), also an instantaneous scintillation signal~(S1). Recently, the XENON1T collaboration has published electronic recoil data that utilizes both S1 and S2 \cite{Aprile:2020tmw}, and last year the analysis of S2-only that extends down to $\sim 200$\,eV. 

\begin{figure*}[tb]
	\includegraphics[width=0.488\textwidth]{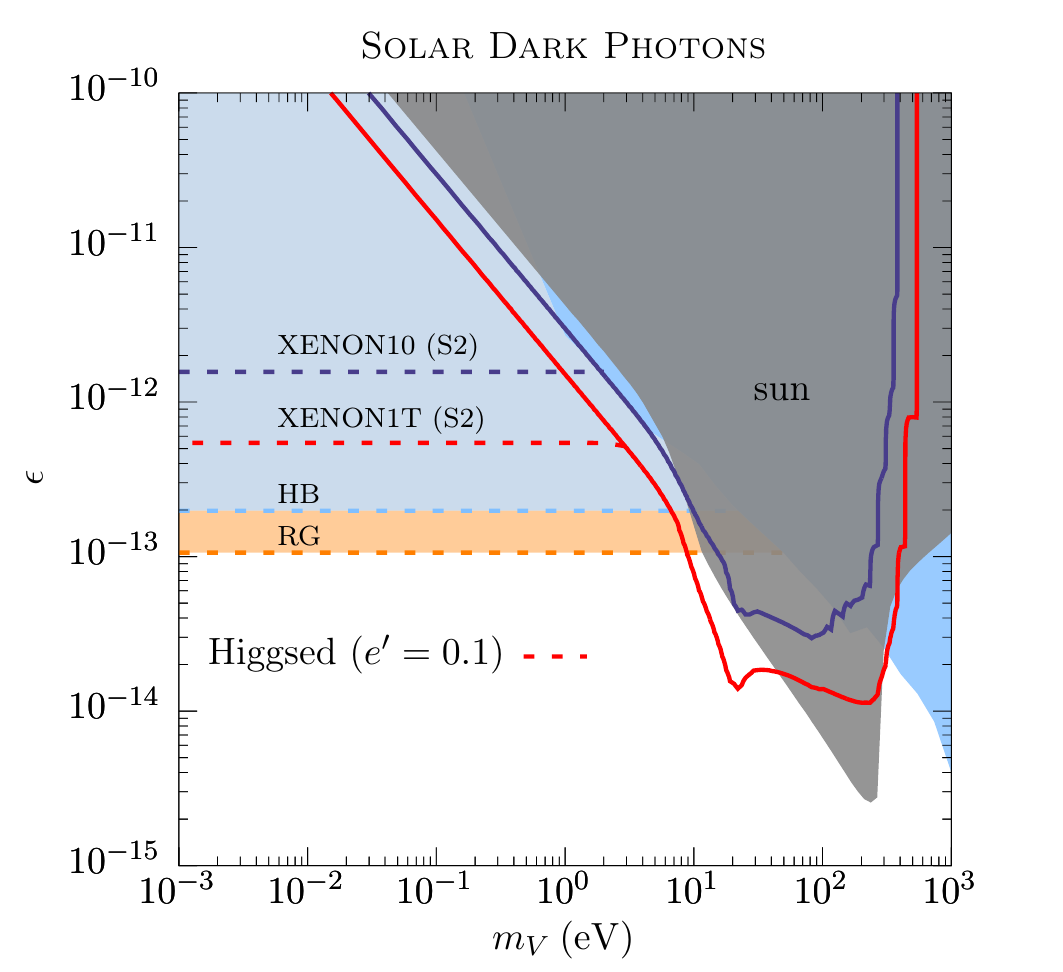}%
	\hfill
	\includegraphics[width=0.48\textwidth]{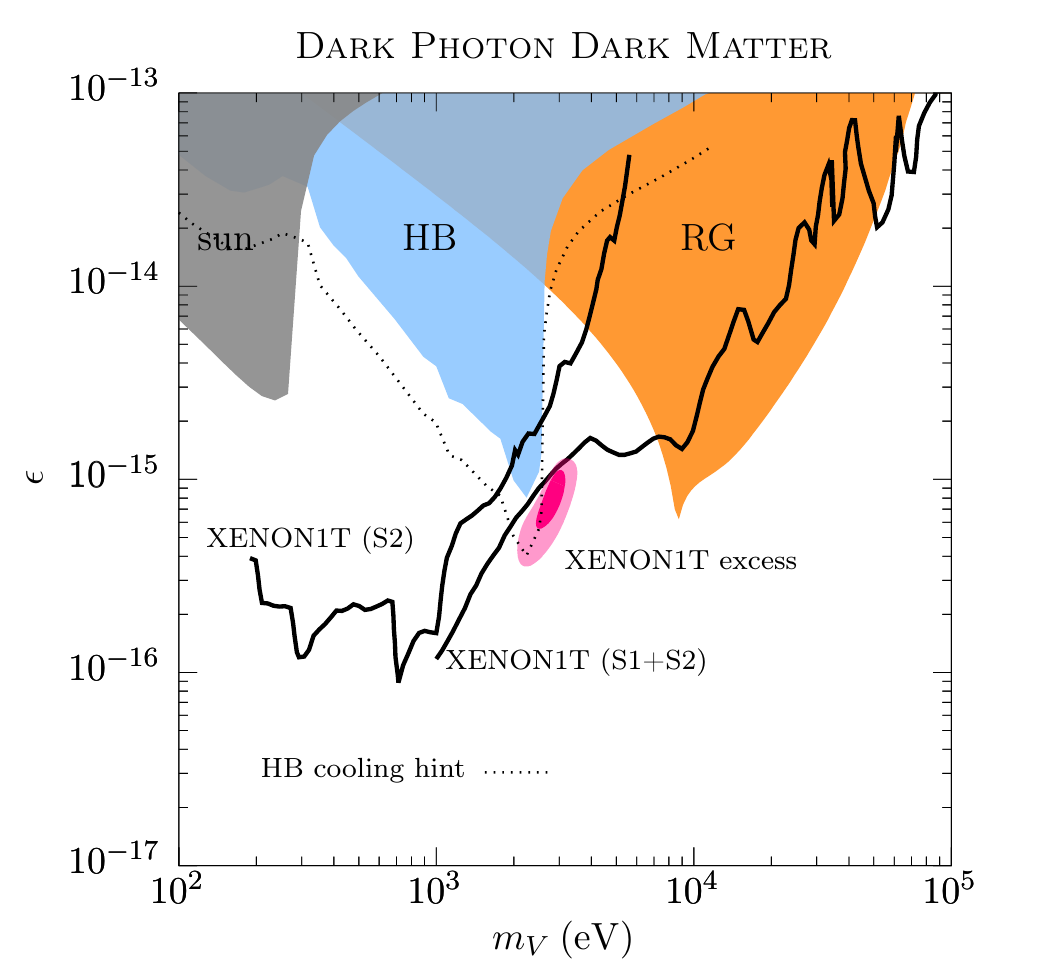}
	\caption{\textit{Left panel:}
	Direct detection constraints at 90\%~C.L.~on solar-generated dark photon fluxes in the parameter space of vector mass~$m_{A'}$ versus kinetic mixing parameter~$\epsilon$. The red (blue) line is derived from the S2-only reported data by XENON1T~\cite{Aprile:2019xxb} (XENON10~\cite{Angle:2011th}). Solid lines apply to a ``hard'' St\"uckelberg mass and dashed lines show how the constraint continues for a ``soft'' Higgsed dark photon mass with $e'=0.1$ and following~\cite{An:2013yua}. Cooling constraints from the sun, and for HB and RG stars as labeled are derived following~\cite{An:2013yfc,An:2014twa}.	\textit{Right panel:} Dark photon dark matter parameter space showing the favored region from a fit to XENON1T data~\cite{Aprile:2020tmw} ($1\sigma$ and $2\sigma$ ellipses). Official limits by the XENON1T collaboration using  S2~\cite{Aprile:2019xxb} and S1+S2~\cite{Aprile:2020tmw} data are shown by the solid black lines as labeled. The HB constraint (and cooling hint, dotted line) are taken from~\cite{Giannotti:2015kwo} and the solar and RG constraints are derived following~\cite{An:2013yfc,An:2014twa}; see the main text for a discussion of the latter bounds.
	}\label{fig:1}
\end{figure*}

The increased sensitivity for solar dark photons comes mainly from the XENON1T S2-only analysis~\cite{Aprile:2019xxb}. 
The probability to observe S2 photo-electrons (PE) following an absorption event with energy deposition $\Delta E$ is given by $   P(S2|  \Delta E  ) = \sum_{n_e, n_e^{\rm surv}}
  P(S2|n_e^{\rm surv})   P(n_e^{\rm surv} | n_e) P(n_e | \langle n_e \rangle )$. Here, $ P(n_e | \langle n_e \rangle ) = \binomial (n_e |  N_Q , f_e ) $ is the binomial probability to create $n_e$ ionized electrons from $N_Q = \Delta E /(13.8~\eV)$~\cite{Dahl:2009nta,Akerib:2016qlr} trials with 
single event probability $f_e = \langle n_e \rangle /N_Q$. The expectation value for the charge yield, $\langle n_e \rangle $, is measured~\cite{Akerib:2017hph} for $\Delta E \geq 190\,\eV$; below we use the model of~\cite{Essig:2012yx} with vanishing recombination probability.  For $ P(n_e^{\rm surv} | n_e)  $ we then assume  that  80\% (100\%) of electrons successfully drift to the gas-liquid interface in XENON1T (XENON10). The resulting scintillation signal is then assumed to be Gaussian, $  P(S2|n_e^{\rm surv}) = \gaussian(S2| g_2 n_{e^{\rm surv}}, \sigma_{S2})  $, with $\sigma_{S2} = 7 \sqrt{n_e^{\rm surv}}$~\cite{Aprile:2013blg} and a gain factor $g_2 = 33\, (27)$~PE/$e^-$ for XENON1T (XENON10). Finally, the experimentally observed rate is then given by folding the theoretical absorption spectrum $dR/d\Delta E$, $dR/dS2 = \epsilon(S2) \int d\Delta E\, P(S2|\Delta E)  dR/d\Delta E $ where $\epsilon(S2)$ is the S2 detection efficiency~\cite{Angle:2011th,Aprile:2019xxb}.

The theoretical energy deposition rate that feeds into the computation of the experimental signal is given by, 
\begin{align}
\label{thspec}
    \frac{dR}{d\Delta E} = \frac{1}{\rho_{\rm Xe} \beta_{A'}} \int_{\Delta E} dE \, \frac{d\phi}{dE} \frac{d\Gamma}{d\Delta E} .
\end{align}
Here $\beta_{A'}$ is the incoming dark photon velocity and $\rho_{\rm Xe} = 3.1\,{\rm g}/\cm^3$ is the representative mass density of liquid xenon. The energy-differential solar fluxes $d\phi/dE$ were computed in~\cite{An:2013yfc, An:2013yua}. In the following we are going to consider both cases, when the dark photon has a ``hard mass'' (St\"uckelberg case) or a ``soft mass'' (Higgsed case). In the former case, the dark photon is strictly absorbed and the entire energy deposited in the detector, \begin{align}
    \frac{d\Gamma}{d\Delta E} = \Gamma_{\rm abs} \delta(E - \Delta E) .
\end{align}
In the latter case, we have the additional process of scattering, $A'  \to h$ or $h\to A'$,  where $h$ is the Higgs particle  associated with the mass-generation of $m_{A'}$ and where an amount of $\Delta E$ is transferred to the atom via the exchange of $A'$; both rates $\Gamma_{\rm abs}$ and $   {d\Gamma}/{d\Delta E}|_{\rm higgsed}$ were derived in~\cite{An:2013yua}; $m_h\sim m_{A'}$ is assumed for the latter.

The new constraints, plotted as a solid red line are shown in Fig.~\ref{fig:1} (left panel). They supersede earlier results based on XENON10 data. Shown in dark 
grey are the solar energy loss bounds based on the requirement that not more than 10\% of the solar luminosity is emitted in the form of dark states, $L_{A'} <0.1L_{\odot}$. (Notice that a more aggressive use 
of different components in the neutrino flux and solar composition has been claimed to strengthen the bounds by an additional factor of a few, to $L_{A'} <0.02L_{\odot}$ \cite{Vinyoles:2015aba}.)
Our new results based on XENON1T data, taken in the scaling regime correspond to the constraint $L_{A'} <0.013L_{\odot}$. It is now clear that the sensitivity of direct detection experiments has overtaken the solar energy loss constraints, and it is uncertain that future progress in understanding the solar composition would be able to overcome this sensitivity gap. The horizontal dashed lines show the 
constraints on the ``Higgsed" dark photon, where the mass comes from spontaneous symmetry breaking in the dark sector, and the emission inside the Sun proceeds via 
$A'h'$ pair production. XENON1T provides a significant improvement over the 
previous limit from XENON10 data, but is still subdominant to the stellar bounds. 

It is also possible to significantly improve constraints on other light vector portal models. For example, some 
anomaly-free combinations of nearly or exactly preserved 
symmetries in the Standard Model can be gauged. The best known examples include the $B-L$ combination, as well as flavor-specific lepton number combinations such as $L_{\mu}-L_{\tau}$. For vectors coupled to the $B-L$ current, the contribution of neutrons in the emission of $V_{B-L}$ 
is small \cite{Hardy:2016kme}, and the energy loss limits 
largely follow the dark photon case. In addition, there are new constraints from coherent forces mediated by $V_{B-L}$. As a result, our limits can be translated via the replacement of $e\epsilon \to g_{B-L}$, and improve the current bounds in the $m_V$ interval from $\sim 10$ to $\sim 100$ eV. 
(Notice that for such small values of the coupling, the decay $V \to \nu\bar \nu $ happens beyond 1 A.U.) Finally, even models that do not have a direct coupling to the electron current, such as $L_\mu-L_\tau$, and gauged 
baryon number with the anomaly cancelled at or above the weak scale, can be probed by XENON1T. This is because at the radiative level, the 
kinetic mixing operator is always induced (and its cancellation can only be arranged via fine-tuning of the UV boundary conditions). Our results can therefore be 
generalized to these models as well.

\section{Absorption of keV-scale dark matter and the XENON1T excess}
\label{sec:absorption}

The most recent electron recoil data from XENON1T~\cite{Aprile:2020tmw}, that uses both S1 and S2 signals, shows a slight excess over the background in the 2-5 keV energy bins. This excess has sparked a series of investigations by different groups addressing whether it can be explained by new physics
\cite{Kannike:2020agf,*Fornal:2020npv,*Boehm:2020ltd,*Harigaya:2020ckz,*Bally:2020yid,*Su:2020zny,*Du:2020ybt,*DiLuzio:2020jjp,*Bell:2020bes,*Chen:2020gcl,*AristizabalSierra:2020edu,*Buch:2020mrg,*Choi:2020udy,*Paz:2020pbc,*Primulando:2020rdk,*Dey:2020sai} or additional background~\cite{Szydagis:2020isq}.

Dark matter composed of light keV mass particles could be a natural explanation for the excess. One has to be mindful of the fact that 
the mass of dark matter is a free parameter, and the local significance of the peak is washed out in the global significance by the ``look elsewhere" effect. 
Nevertheless, the absorption of sub-5-keV dark photons that saturate or nearly saturate the local dark matter energy density is an interesting case to consider. Owing to the large number densities, the absorption of dark photon dark matter provides sensitivity to even smaller couplings than solar emission~\cite{An:2014twa}.

 To test the dark photon dark matter signal hypothesis, we fit the theoretical absorption spectrum,
\begin{equation}
\label{dpdm}
 \frac{dR}{d\Delta E} = \frac{1}{\rho_{\rm Xe}} \frac{\rho_{\rm dm}}{m_{A'}} \Gamma_{\rm abs} \delta(\Delta E - m_{A'}) ,
\end{equation}
to the reported  electron recoil spectrum in~\cite{Aprile:2020tmw}. Here, $\rho_{\rm dm} \simeq 0.3\,\GeV/\cm^3$ is the local dark matter density. Equation~\eqref{dpdm} is obtained from~\eqref{thspec} with $d\phi/dE = (\rho_{\rm dm}/m_{A'}) \beta_{A'} \delta(E-\Delta m) $ and to which we subsequently apply detection efficiency and energy resolution as reported in~\cite{Aprile:2020tmw} and~\cite{XENON:2019dti}. Adding the background as given in~\cite{Aprile:2020tmw}, the region of interest in the $\epsilon$-$m_{A'}$ parameter space is then obtained from a Poisson log-likelihood test statistic and $68\%$ ($95\%$) confidence regions are constructed from a worsening by $\Delta\chi^2 = 2.30 (5.99)$ from the $\chi^2$-value of the best fit point.

The inferred regions of interest given by the pink-shaded ellipses in the right panel of Fig.~\ref{fig:1} show that the region of parameter space that can give an elevated counting rate consistent with the XENON1T excess lies generally below $\epsilon= 10^{-15}$, and just outside the region excluded by energy loss constraints from horizontal branch stars. Owing to the absence of two-photon decays, in that mass range this model also satisfies observational X-ray data~\cite{Pospelov:2008jk,An:2014twa}. We note in passing that we update the red giant constraint from~\cite{An:2014twa} using a revised average helium core density of $\rho = 2\times 10^5\,{\rm g/\cm^3}$ instead of $\rho =  10^6\,{\rm g/\cm^3}$ with a correspondingly smaller core-averaged plasma frequency of $\omega_p = 8.6\,\keV$~\cite{Chu:2019rok}.

\begin{figure*}[tb]
	\includegraphics[width=0.5\textwidth]{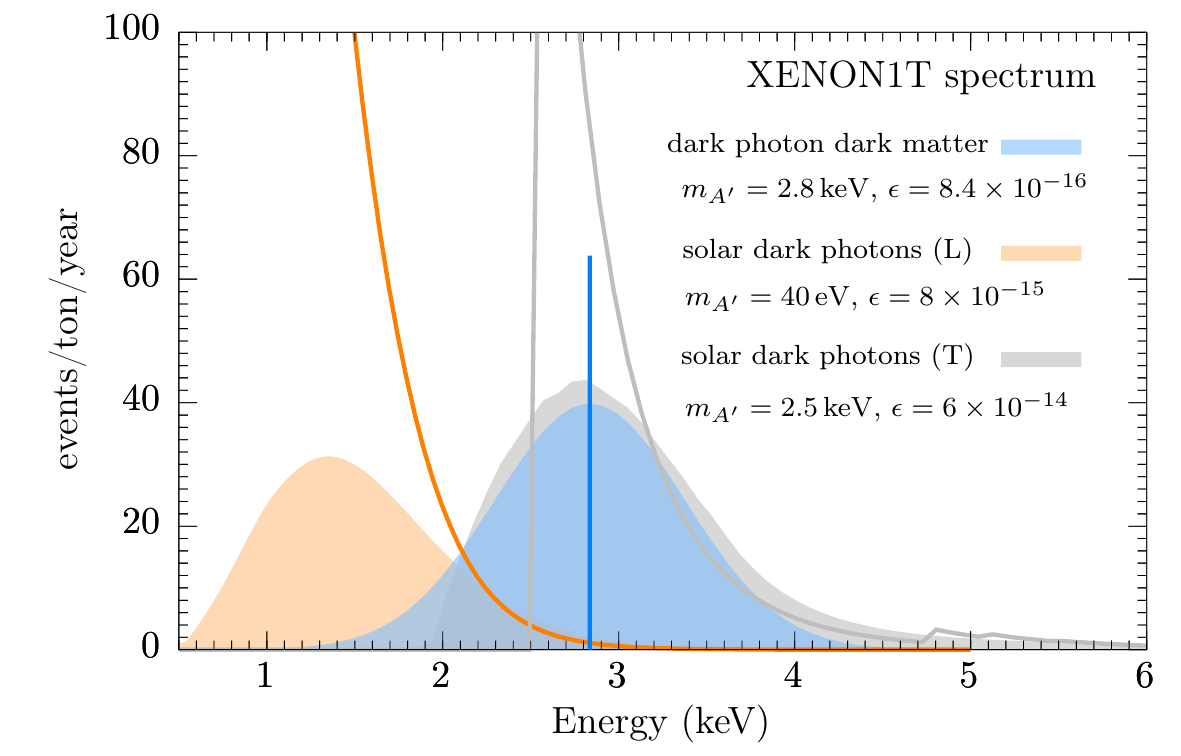}%
	\includegraphics[width=0.5\textwidth]{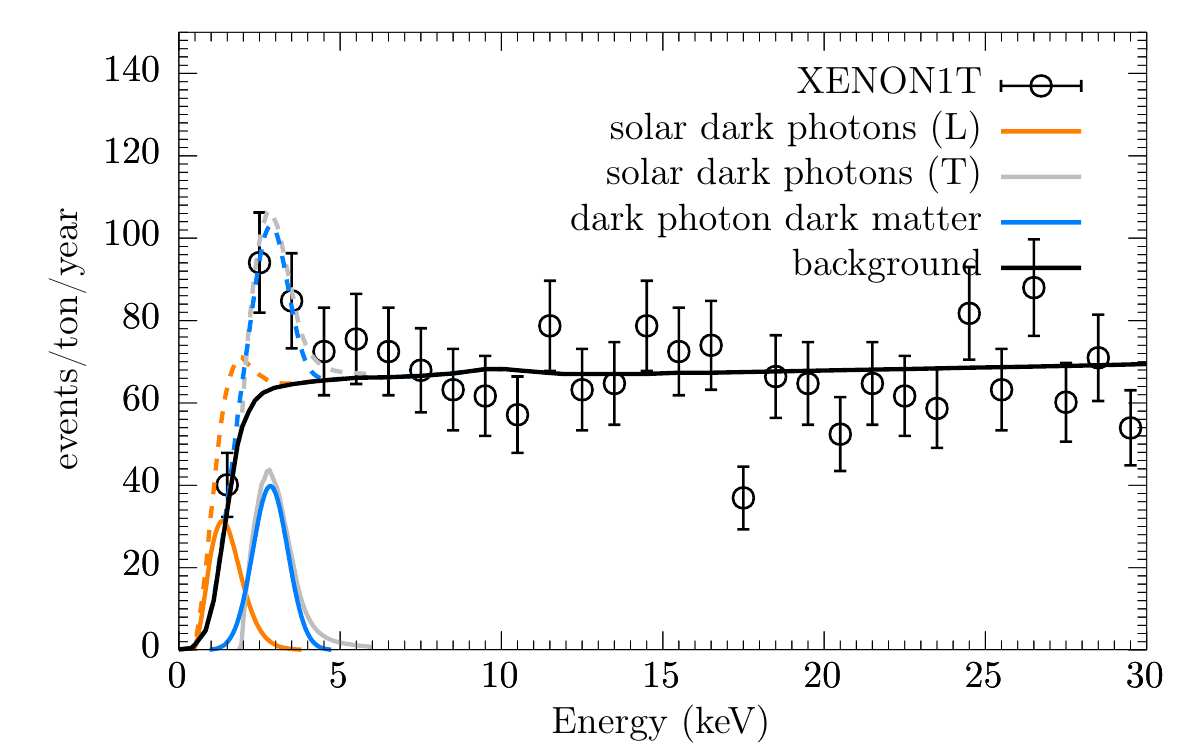}
		\caption{\textit{Left panel:} The best dark photon dark matter fit and exemplary parameter point for a solar generated longitduinal (L) and transverse (T) dark photon flux. The lines show the theoretical electron recoil spectra, the shaded regions show the spectra with detection efficiency and resolution folded in. 
	\textit{Right panel:}
	XENON1T-reported data on recoil events with an S1 signal component~\cite{Aprile:2020tmw}. The reported background prediction is shown by the black line. The blue solid (dashed) line shows the signal (signal+background) for absorption of dark photon dark matter. The orange and gray lines show exemplary signals for a solar dark photon-generated L and T flux, respectively; $m_{A'}$ and $\epsilon$ as per left panel. Although transverse solar dark photons with 2.5~\keV\ mass provide a good fit to the data, the strength of $\epsilon$ is excluded (see previous figure).}\label{fig:2}
\end{figure*}

Figure~\ref{fig:2} provides an example of the expected signal for the choice of parameters $m_{A'}=2.8$~keV and $\epsilon =8.4\times 10^{-16}$, corresponding to the best fit point. The goodness-of-fit using Pearson's $\chi^2$ for signal plus background is 
$\chi^2/\text{d.o.f.} = 35.5/27$ ($p=12.7\%$) compared to the background-only hypothesis of $\chi^2/\text{d.o.f.} = 46.3/29$ ($p=2.2\%$). For comparison, we also include the solar dark photon contribution for the exemplary parameter point $m_{A'}= 40\,\eV$ and $\epsilon = 8\times 10^{-15}$. The flux is overwhelmingly in longitudinal dark photons. The spectrum generally peaks at or below the threshold, and throughout the considered parameter region, no noticeable improvement on the goodness-of-fit is observed as the signal will overshoot the first data-point without filling the subsequent anomalous bins. 
A good fit to the XENON1T excess from a solar dark photon flux can only be obtained with a sufficiently heavy mass, $m_{A'}\simeq 2.5\,\keV$ and $\epsilon \simeq 6\times 10^{-14}$, for which the flux is in transverse modes. However, the favored parameter range then finds itself inside the astrophysically excluded region, see Fig.~\ref{fig:1}.

\section{Discussion} 
\label{sec:discussion}

We have updated the existing bounds on solar dark photons.
The XENON1T experiment, using the S2-only signal \cite{Aprile:2019xxb}, provides the 
most stringent constraint to date, improving on the ``optimized"
solar limits in \cite{Vinyoles:2015aba}. Our strengthened bounds bear implications for the 
requisite sensitivity of proposed dark matter experiments that seek to use layered optical materials for the absorption of (sub-)eV dark photon dark matter~\cite{Baryakhtar:2018doz}. 
It is very unlikely, due to the much softer shape of the spectrum, that solar dark photons with $m_{A'}\leq 1\,\keV$ could cause the XENON1T excess in the 2-5 keV window; for  $m_{A'}\geq 1\,\keV$ the required value of $\epsilon$ is excluded by stellar cooling.
At the same time, absorption of light dark matter does appear to be a viable, if ad-hoc, explanation for the current excess (see the most recent works \cite{Takahashi:2020bpq,Alonso-Alvarez:2020cdv}). In agreement with~\cite{Alonso-Alvarez:2020cdv}, we find that dark photon dark matter provides an acceptable fit to the current data while satisfying stellar bounds. 

It is natural to ask whether dark photon dark matter provides the most suitable vector portal candidate for the explanation of the excess. At the level of dimension 4 interactions, it is clear that one can generalize dark photons to other vector dark matter candidates. However, a coupling to the $B-L$ and $L_\mu-L_\tau$ currents will not work, as 
the lifetimes for such states (given the XENON1T-motivated range of couplings) are expected to be shorter than the lifetime of the Universe due to $V\to \nu\bar \nu $ decays. (It may be interesting to explore whether the emission of keV mass dark vectors into a Solar bound orbit \cite{VanTilburg:2020jvl} could potentially furnish a sufficient number density to generate a signal.)
On the other hand, the coupling of a vector $V$ to the baryonic current, 
$V_\mu \sum_q (g_B/3)\bar q \gamma_\mu q $, may be used due to its radiative mixing with electromagnetic current, essentially reducing this case to that of the dark photon. Other sensitive constraints on vectors coupled to baryon number, based on the anomalous production of longitudinal modes
\cite{Dror:2017ehi,Dror:2017nsg} cannot reach the level of $g_B/m_V \propto 10^{-15}{\rm keV}^{-1}$ probed by the XENON1T data. 
A slight conceptual advantage of kinetically-mixed dark photons over other models with an exceedingly 
small gauge coupling constant is that it is easier to 
model the emergence of a small mixing angle. For example, mechanisms for radiatively generating $\epsilon$ include multi-loop and even gravitational loop mediation 
\cite{Dunsky:2019api,Koren:2019iuv,Gherghetta:2019coi}, so that values of $\epsilon$ in the $O(10^{-15})$ range may not necessarily be exotic.

We can also consider higher dimensional interaction channels, where it is natural to focus on axion-like particles. 
As was shown in \cite{Takahashi:2020bpq} (see also \cite{Pospelov:2008jk}), axion-like  particles in the few keV mass range, that are derivatively coupled to the electron current, can induce a similar signal 
while being protected from rapid $\gamma\gamma$ decays. It is clear that other derivatively-coupled fields may be used as well. (The keV-scale mass precludes an axion-like particle from being a QCD axion, and therefore there is no particular advantage over other models with derivative couplings.) A new vector particle can also couple to the ``dark" magnetic or electric dipole moment operators \cite{Dobrescu:2004wz}. Taking the magnetic moment case, 
${\cal L}= (\mu/2) V_{\mu\nu} \bar e \sigma_{\mu\nu}e$, and saturating the dark matter abundance with $V_\mu$, will lead to a similar experimental signature: ionization of atoms induced by the absorption of $V_\mu$. 
The ionization cross section due to 
$\mu$ (taking the leading order in $Z\alpha$, and the simple approach to atomic physics in \cite{Pospelov:2008jk}) 
can be related to the photoelectric effect as 
$\sigma_{V} v /\sigma_{\rm photo} c \simeq (\mu m_V)^2/ (2\pi\alpha)$, 
which up to a small numerical difference is equivalent to an axion-like particle with the
$g_{aee}/m_e \to \mu$ substitution. It is then clear that  $\mu$ in the range of $10^{-11}-10^{-10} \,{\rm GeV}^{-1}$ and the same mass as above will be equally suitable in modeling the excess counts.

As a final remark, we can also comment on another class of models where {\em partial} de-excitation of dark matter $X_2\to X_1$ in the interaction with electrons and/or nuclei may occur (see {\em e.g.} \cite{Batell:2009vb,Bernal:2017mqb}). 
Choosing the mass splitting $\Delta m$ in the ``right" energy range, one can easily arrange for the de-excitation process to 
deposit a few keV of energy to electrons, and 
create an excess. This model suffers from the same degree of 
arbitrariness as the models discussed above involving absorption of keV dark matter. The difference 
will be in the mass of dark matter that can now be chosen to be above $\sim 10$\,MeV, so that cosmological and stellar and supernova energy loss constraints may be less stringent in this scenario. 

In conclusion, the arrival of ton-scale xenon-based underground detectors that achieve electronic background counts as low as $O(10-100){\rm /yr/ton}$ heralds a new era, enabling the experimental study of new phenomena that deposit very small amounts of energy (keV and sub-keV) per interaction. This technology, and its future evolution to multi-ton scale experiments will allow the expanding study of both neutrinos and of novel particles and interactions that are otherwise invisible to other probes.

\paragraph*{Acknowledgements.} HA is supported by NSFC under Grant No.~11975134, the National Key Research and Development Program of China under Grant No.~2017YFA0402204 and Tsinghua University Initiative Scientific Research Program. MP is supported in part by 
U.S. Department of Energy (Grant No. DE-SC0011842). JP is supported by the New Frontiers Program by the Austrian Academy of Sciences and by the Austrian Science Fund (FWF) Grant No.~FG~1. AR is supported in part by NSERC, Canada. 

\bibliographystyle{apsrev4-1}
%\bibliography{refs}{}

\end{document}